\documentclass[preprint,12pt]{elsarticle}
\biboptions{sort&compress}
\usepackage[utf8]{inputenc}
\usepackage[utf8]{inputenc} 
\usepackage[T1]{fontenc}
\usepackage{textcomp}       
\usepackage[numbers]{natbib}
\usepackage[version=3]{mhchem} 
\usepackage{gensymb}
\usepackage{tablefootnote}
\usepackage{threeparttable}
\usepackage{subcaption}

\usepackage{anyfontsize}
\usepackage{textcomp}
\usepackage[dvipsnames]{xcolor}

\usepackage{amssymb}

\begin{document}

\begin{frontmatter}



\title{A Molecular Beam Study of the (0,0)$A^2\Pi \leftarrow X^2\Sigma^+$ Band of CaF Isotopologues}


\author[inst1]{Termeh Bashiri}
\author[inst2]{Phelan Yu}
\author[inst2]{Nicholas R. Hutzler}
\author[inst2,inst3]{Timothy C. Steimle}
\author[inst4]{Carlos Abad}
\author[inst1]{Mitchio Okumura}

\affiliation[inst1]{organization={Division of Chemistry and Chemical Engineering},
            addressline={California Institute of Technology, 1200 E California Blvd}, 
            city={Pasadena},
            state={CA},
            postcode={91125},
            country={USA}}
\affiliation[inst2]{organization={Division of Physics, Math, and Astronomy},
            addressline={California Institute of Technology, 1200 E California Blvd}, 
            city={Pasadena},
            state={CA},
            postcode={91125},
            country={USA}}
\affiliation[inst3]{organization={School of Molecular Sciences},
            addressline={Arizona State University, 1151 S Forest Ave}, 
            city={Tempe},
            state={AZ},
            postcode={85287}, 
            country={USA}}
\affiliation[inst4]{organization={Bundesanstalt für Materialforschung und -prüfung (BAM)},
            addressline={Richard-Willstätter-Straße 11}, 
            postcode={12489},
            city={Berlin},
            country={Germany}}

\begin{abstract}
This study presents an experimental determination of spectroscopic parameters for the less-abundant isotopologues \ce{^42CaF} and \ce{^44CaF}, alongside \ce{^40CaF}, by high resolution laser-induced fluorescence spectroscopy in a skimmed free jet expansion. We recorded spectra near the natural linewidth limit and derived spectroscopic constants for both $X^2\Sigma^+$ and $A^2\Pi$ states, including the fine and \ce{^19F} magnetic hyperfine parameters. We also estimated the \textit{r}(\ce{^44Ca}/\ce{^40Ca}) isotope amount ratio, demonstrating the potential use of optical spectroscopy for isotope analysis. 

\end{abstract}




\end{frontmatter}

\section{Introduction}
The measurement of calcium isotopes plays pivotal roles in the investigation of biological, environmental, and geological processes. Calcium isotope amount ratios r($^{\text{heavy}}$Ca/$^{\text{light}}$Ca), hereafter isotope ratios, have been proposed as a biomarker for the diagnosis of bone cancer and chronic kidney disease \cite{channon_using_2015,tacail_new_2020,shroff_naturally_2022}. They also serve as proxies for the paleo-environment due to the coupling between the geochemical calcium and carbon cycles \cite{kovacs_calcium_2022}. However, conventional isotope ratio analysis methods, such as thermal ionization mass spectrometry (TIMS) \cite{rturnlund_isotope_1993, depaolo_calcium_2004} and multi-collector inductively coupled plasma mass spectrometry (MC-ICP-MS) \cite{he_practical_2019}, face significant challenges for Ca isotope analysis due to isobaric interferences arising from overlapping isotope lines, particularly \ce{^40Ar+}. These interferences limit calcium isotope analysis to expensive and high-resolution instruments, which still preclude the detection of the most abundant isotope, \ce{^40Ca}. This highlights the inherent limitations of mass spectrometry for the analysis of calcium isotopes. Although high-resolution MC-ICP-MS instruments and rigorous sample preparation techniques, including matrix extraction, can mitigate these issues, they still rely on calibration against certified reference materials with imperfectly known isotope compositions. 

High-resolution spectroscopy presents a promising alternative to mass spectrometric methods for isotopic analysis \cite{abad_determination_2017, winckelmann_high-resolution_2021, morcillo_lithium_2024}. However, atomic spectroscopy can be challenging due to the small isotope shifts observed in atomic spectra. The spectral shift between \ce{^40Ca} and \ce{^44Ca} is 996.2~MHz for the visible inter-combination transition $4s4p$~$^3P_1$~$\leftarrow$ $4s^2$~$^1S_0$ near 675.5~nm \cite{bergmann_nuclear_1980}, which is below the 2.2~GHz Doppler broadened linewidth \cite{miyabe_doppler-free_2023}. Sub-Doppler techniques such as saturation-dip or two-photon absorption spectroscopy, needed for resolving atomic calcium isotope lines, require high-intensity lasers and complex experimental setups to resolve small isotope shifts, making them impractical for routine sample analysis in real-world conditions. Additionally, these methods are problematic due to their non-linear response.

Diatomic molecules have larger isotopic shifts in their spectra due to the large mass effects on the rotational and vibrational energies of the molecule, making them an attractive alternative to atoms \cite{tzanatta_calcium_2019}. Recently, Abad and coworkers (\cite{kowalewska_feasibility_2023,abad_zirconium_2018}) have shown that the isotopic compositions of atoms can be determined by electronic absorption spectroscopy of their diatomic analogues. For alkaline earth metals, these experiments take advantage of the strong, well-resolved optical transitions of radical diatomics, e.g. SrF \cite{bazo_high-resolution_2022}. Dried samples of salts are volatilized in a graphite oven, and spectra are recorded in a modified Atomic Absorption Spectrometer. Their approach provides sensitivities and precisions similar to that achievable by Isotope Ratio Mass Spectrometry (IRMS). 

The open-shell diatomic CaF provides an excellent candidate for isotopic analysis of Ca. CaF has a strong electronic transition (A$-$X) with an origin at 606 nm. Additionally, given that fluorine is monoisotopic, there are no further splittings in the spectra due to mixed isotopes. 

Although the spectroscopy of the most abundant isotopologue, \ce{^40CaF}, has been studied extensively \cite{anderson_millimeter-wave_1994, kaledin_analysis_1999, field_continuous_1975, devlin_measurements_2015, lavy_line_2025}, there have been no previous experimental investigations of the less abundant isotopologues. Hou and Bernath \cite{hou_line_2018} have predicted the vibrational $G_v$, and rotational $B_v$, parameters for the $X^2\Sigma^+$ ($\nu=0-10$) levels of $\ce{^42CaF}$, $\ce{^43CaF}$, $\ce{^44CaF}$, $\ce{^46CaF}$, and $\ce{^48CaF}$. These parameters were obtained using the experimental vibration-rotation \cite{charron_high-resolution_1995} and pure rotational data \cite{anderson_millimeter-wave_1994} for $\ce{^40CaF}$ to generate an extended Morse oscillator (EMO) potential. The resulting EMO potential was used as input for a direct solution of the Schrödinger equation for nuclear motion using LeRoy’s computer program LEVEL \cite{le_roy_level_2017}. The calculated ro-vibrational levels from the LEVEL program were then modeled using a simple analytical expressions (i.e. E($\nu,J$)= $G_v$($\nu+1/2$) + $B_vJ(J+1)$) to predict vibrational and rotational parameters. The electron spin-rotation and \ce{^19F} hyperfine interactions were ignored (i.e. the $X^2\Sigma^+$ state of CaF was treated as $\Sigma^+$ state). More recently, the same group performed new optical spectroscopic measurements at a spectral resolution of 0.05 cm$^{-1}$ \cite{lavy_line_2025}. The results were combined with existing data to generate an extensive characterization of the $X^2\Sigma^+$, $A^2\Pi$, and $B^2\Sigma^+$ states. The analysis was limited to the main \ce{^40CaF} isotopologues and was used to generate a line list of visible transitions of a hot sample (rotational temperature of 2500 K and vibrational temperature of 4000 K) with a Gaussian linewidth of 0.08 cm$^{-1}$ \cite{lavy_line_2025}.

The objective of the present work is to experimentally determine spectroscopic parameters for \ce{^40CaF}, \ce{^42CaF}, and \ce{^44CaF} in the $A^2\Pi$ -- $X^2\Sigma^+$ band system. Here we present the analysis of the (0,0)$A^2\Pi$ -- $X^2\Sigma^+$ band recorded at near natural linewidth limit and with sufficient sensitivity to detect the even \ce{^40CaF}, \ce{^42CaF}, \ce{^44CaF} isotopologues. We determine spectroscopic constants, including fine structure parameters, and compare to the
EMO-predicted parameters. Furthermore, a comparison of the determined fine and \ce{^19F} magnetic hyperfine parameters for the $A^2\Pi$~($\nu$=0) and $X^2\Sigma^+$~($\nu$=0) states for the three isotopologues is discussed. The spectral intensities of multiple lines were used to make an estimate of the  $\textit{r}$(\ce{^44Ca}/\ce{^40Ca}) isotopic ratio.

\section{Experimental}

The pulsed molecular beam spectrometer used in the present work is nearly identical to that used in the previous optical Zeeman study of the (0,0)$A^2\Pi-X^2\Sigma^+$ and (0,0)$B^2\Sigma^+-X^2\Sigma^+$  bands of \ce{^40CaF} \cite{devlin_measurements_2015}. Briefly, a cold (T$_{rot}$ = 20~K) pulsed molecular beam sample of CaF was generated by skimming the output of a laser ablated calcium sample (ESPI Metals, 99.5$\%$ purity) entrained, and reacting with a supersonic expansion of 5$\%$ \ce{SF6}/95$\%$ Ar mixture. The molecular beam was probed at approximately 75~cm from the source with a low intensity ($\sim$ 1~mW and 2~mm diameter), tunable, narrow linewidth beam (\textless 1MHz) of a continuous-wave dye laser. The resulting on-resonance laser induced fluorescence (LIF) viewed through a 620$\pm$10~nm bandpass filter was detected using a cooled photomultiplier tube and the signal was processed using photon counting techniques. Typically, the counts from 20 pulsed samples were summed at a given laser excitation wavelength. The laser wavelength was calibrated using a commercial wavemeter combined with a co-recorded sub-Doppler \ce{I2} standard \cite{salumbides_improved_2008}.

The collected spectra were analyzed by fitting to a linear molecule effective Hamiltonian using the PGOPHER  software suite \cite{western_automatic_2017,western_automatic_2019,western_pgopher_2017}. Spectroscopic parameters were then extracted from the resulting fit, which are provided and discussed in the next section. 

\section{Results and discussion}

\subsection{Spectroscopic Parameters}

\begin{figure}[ht]
    \centering
    \includegraphics[width=0.925\linewidth]{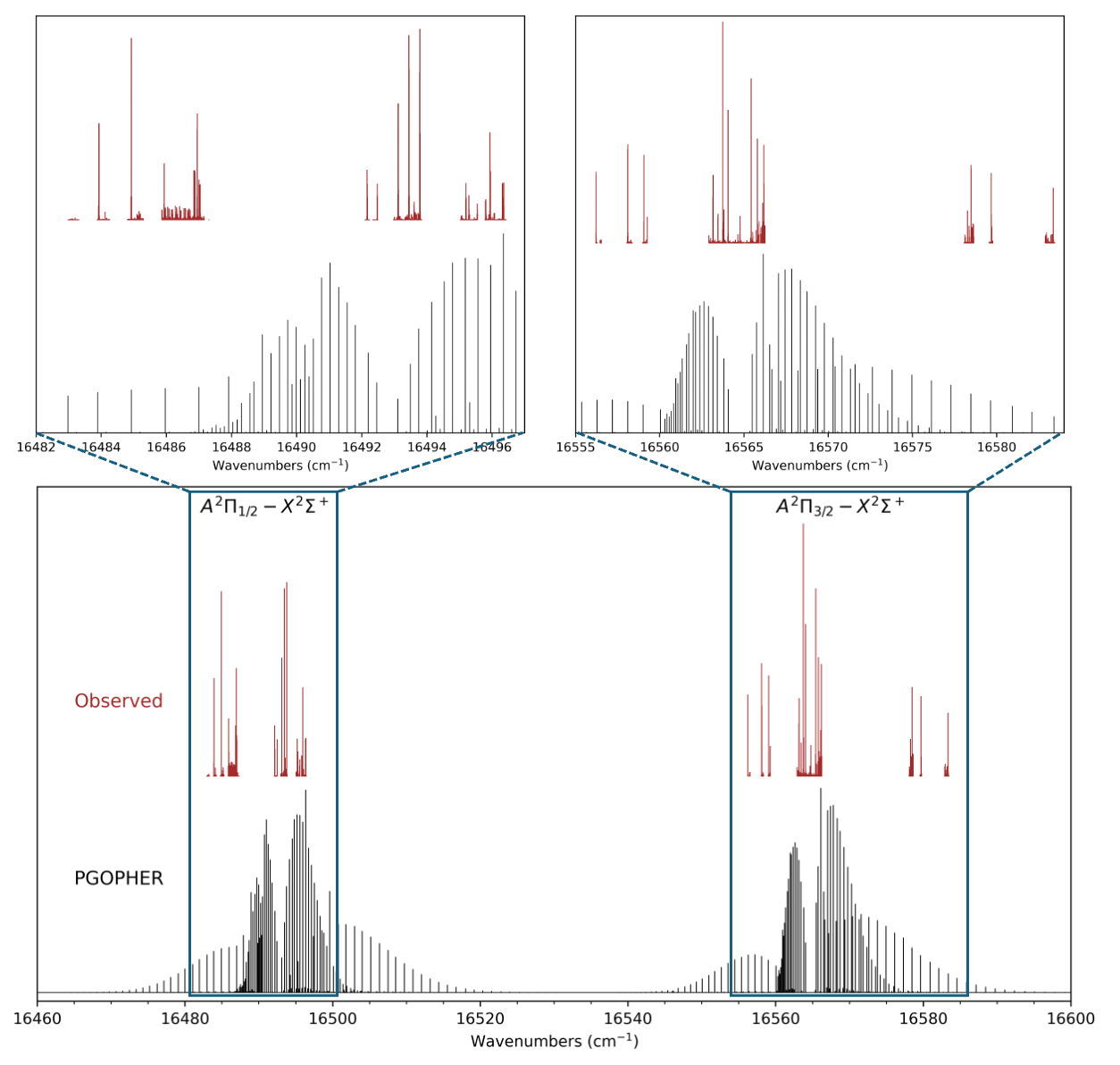}
    \caption{The experimentally observed spectrum (in red) and the simulated spectrum (in black) showing the (0,0)$A^2\Pi$~--~$X^2\Sigma^+$ and (0,0)$A^2\Pi$~--~$X^2\Sigma^+$ bands.}
    \label{fig:whole_spectrum}
\end{figure}

The experimental and simulated spectra are shown in Figure \ref{fig:whole_spectrum}. The input data set consisted of 144, 21, and 72 lines for the \ce{^40CaF}, \ce{^42CaF}, and \ce{^44CaF} isotopologues, respectively (Table~\ref{tbl:properties}). Since our primary goal was to investigate the transitions of less-abundant isotopologues, we targeted specific regions of the spectrum where \ce{^42CaF} and \ce{^44CaF} lines were not heavily overlapped with \ce{^40CaF} lines. As a result, the full $A^2\Pi$~--~$X^2\Sigma^+$ band was not recorded. The unblended spectra features have a full width at half maximum of approximately 40 MHz. 

\begin{table}[ht]
\centering
\caption[]{The natural abundance, number of observed lines, and signal-to-noise ratio of the CaF isotopologues.}\label{tbl:properties}
\small{
\begin{threeparttable}[b]
\centering
\begin{tabular}{llll}\hline
  & Abundance \tnote{a} & Number of observed lines & Signal-to-noise ratio \\ \hline
\ce{^40CaF}  & 96.941(6)$\%$ & 144 & 1270 \\ 
\ce{^42CaF}  & 0.647(3)$\%$ & 21 & 11 \\ 
\ce{^44CaF}  & 2.086(4)$\%$ & 72 & 22 \\ 
\hline\hline
\end{tabular}
\begin{tablenotes}
    \item [a] Calcium isotopic composition of NIST SRM 915a calcium carbonate \cite{coplen_compilation_2002}
\end{tablenotes}
\end{threeparttable}}
\end{table}

The spectroscopic parameters determined for \ce{^40CaF}, \ce{^42CaF}, and \ce{^44CaF} are summarized in Tables \ref{tbl:40caf}, \ref{tbl:42caf}, and \ref{tbl:44caf}, respectively. Lines of $^{40}$CaF, $^{42}$CaF, and $^{44}$CaF were fitted separately. The ground state parameters of $^{40}$CaF were held fixed to the values provided in ref. \cite{anderson_millimeter-wave_1994}. The excited state $^{40}$CaF spectroscopic parameters were compared with those obtained by Devlin et al. \citep{devlin_measurements_2015}, while the ground state parameters $B$ and $D$ for \ce{^42CaF} and \ce{^44CaF} were compared to those calculated by Hou and Bernath \citep{hou_line_2018}. Mass-scaled Dunham coefficients were calculated for $^{42}$CaF and $^{44}$CaF isotopologues by using the experimental parameters for $^{40}$CaF determined in this work. Given the small number of observed lines for the less-abundant isotopologues, centrifugal distortion constants in both states were fixed to the mass-scaled coefficients. Furthermore, $\gamma_D$ was fixed for the three isotopologues to the value provided by Devlin et al. \citep{devlin_measurements_2015}.

\begin{table}[ht] 
\caption[]{The determined spectroscopic parameters for the $^{40}$CaF (0,0) $A^2\Pi$ -- $X^2\Sigma^+$ band system.}\label{tbl:40caf}
\begin{threeparttable}[b]
\centering
\begin{tabular}{llll}\hline
\multicolumn{4}{c}{X$^2\Sigma^+$ $(\nu=0)$ \tnote{a}}\\ \hline
Parameters \tnote{b} & This work \tnote{c} & \begin{tabular}[c]{@{}l@{}}Previous exp. \cite{anderson_millimeter-wave_1994}\\ Anderson et al.\end{tabular} & \begin{tabular}[c]{@{}l@{}}Previous exp. \cite{lavy_line_2025}\\ Lavy et al.\end{tabular} \\ \hline
$B$  & 0.34248822 & 0.34248822(8) & 0.34248811(10)  \\
$D\times10^7$  & 4.6899 & 4.6899(53) & 4.68683(52)   \\
$\gamma$   & 0.0013138 &  0.0013138(28) & 0.00131747(59)  \\
$\gamma_D\times10^8$ & 1.668 &  1.668 & 6.5(120)  \\ \hline
 \multicolumn{4}{c}{A$^2\Pi$ $(\nu=0)$} \\ \hline
 Parameters \tnote{b} & This work & \begin{tabular}[c]{@{}l@{}}Previous exp. \cite{devlin_measurements_2015}\\ Devlin et al.\end{tabular} & \begin{tabular}[c]{@{}l@{}}Previous exp. \cite{lavy_line_2025}\\ Lavy et al.\end{tabular}\\ \hline
 $A$ & 72.61743(38)  &  72.61743(47) & 72.6135(56)    \\
 $B$ & 0.3473884(20) &  0.347395(11) & 0.3473918(26)  \\
 $D\times10^7$ & 4.748(32)  & 4.809 & 4.8017(36)  \\
 $A_D\times10^4$ & $-$1.710(17)  & $-$1.49(23) & $-$1.744(58) \\
 $p + 2q$  & $-$0.04509(11) &$-$0.0452(12) & $-$0.045050(28) \\
 $q$ & $-$0.000298(19) &  $-$0.0002916(16)  &  $-$0.000280(13) \\
 $T_0$ & 16528.75459(20)  &16529.1018(24) & 16528.7473(28) \\ \hline\hline
\end{tabular}
\begin{tablenotes}\small
    \item [a] The ground state parameters have been fixed to the values provided by Ref.~\citenum{anderson_millimeter-wave_1994}.
    \item [b] All units are in wavenumbers (cm$^{-1}$). The numbers in parentheses represent 2$\sigma$ error estimate in the last quoted decimal point.
    \item [c] The hyperfine and nuclear spin-rotation parameters were held fixed to those found in Ref.~\citenum{anderson_millimeter-wave_1994}: $b_F$ = 0.0037192 cm$^{-1}$, $c$ = 0.001108 cm$^{-1}$, $C_I$ = 9.673$\times$10$^{-7}$ cm$^{-1}$.
\end{tablenotes}
\end{threeparttable}
\end{table}

\begin{table}[ht]
\caption[]{The determined spectroscopic parameters for the $^{42}$CaF (0,0) $A^2\Pi$ -- $X^2\Sigma^+$ band system.}\label{tbl:42caf}
\begin{threeparttable}[b]
\centering
\begin{tabular}{llll}\hline
\multicolumn{4}{c}{X$^2\Sigma^+$ $(\nu=0)$}\\ \hline
Parameters \tnote{a} & This work \tnote{b} & \begin{tabular}[c]{@{}l@{}}Mass-scaled $^{40}$CaF\\ Dunham coefficients\end{tabular} & \begin{tabular}[c]{@{}l@{}}EMO computed\\ values\end{tabular} \cite{hou_line_2018} \\ \hline
$B$  &  0.337379(56) & 0.33723618 & 0.33724764 \\
$D\times10^7$ \tnote{c}  & 4.547 & 4.547  & 4.546  \\
$\gamma$   & 0.00127(24) & 0.0012738  & \\
$\gamma_D\times10^8$ & 1.668 (fixed) & & \\ \hline
 \multicolumn{4}{c}{A$^2\Pi$ $(\nu=0)$} \\ \hline
 Parameters \tnote{a} & This work & \begin{tabular}[c]{@{}l@{}}Mass-scaled $^{40}$CaF\\ Dunham coefficients\end{tabular}  \\ \hline
 $A$ & 72.6121(17)  & &     \\
 $B$ & 0.342364(60) & 0.34206122 \\
 $D\times10^7$ \tnote{c} & 4.603 & 4.603 \\
 $A_D\times10^4$ & $-$0.91(50) & & \\
 $p + 2q$  & $-$0.0460(39) & &\\
 $q$ & $-$0.00044(54) & &  \\
 $T_0$ & 16528.74266(76) & \\ \hline\hline
\end{tabular}
\begin{tablenotes}\small
    \item [a] All units are in wavenumbers (cm$^{-1}$). The numbers in parentheses represent 2$\sigma$ error estimate in the last quoted decimal point.
    \item [b] The hyperfine and nuclear spin-rotation parameters were held fixed to those found in Ref.~\citenum{anderson_millimeter-wave_1994}: $b_F$ = 0.0037192 cm$^{-1}$, $c$ = 0.001108 cm$^{-1}$, $C_I$ = 9.673$\times$10$^{-7}$ cm$^{-1}$.
    \item [c] Centrifugal distortion constants $D$’’ and $D$’ kept fixed to the mass-scaled $^{40}$CaF Dunham coefficient values. 
\end{tablenotes}
\end{threeparttable}
\end{table}

\begin{table}[h!]
\caption[]{The determined spectroscopic parameters for the $^{44}$CaF (0,0) $A^2\Pi$ -- $X^2\Sigma^+$ band system.}\label{tbl:44caf}
\begin{threeparttable}[b]
\begin{tabular}{llll}\hline
\multicolumn{4}{c}{X$^2\Sigma^+$ $(\nu=0)$}\\ \hline
Parameters \tnote{a} & This work \tnote{b} & \begin{tabular}[c]{@{}l@{}}Mass-scaled $^{40}$CaF\\ Dunham coefficients\end{tabular} & \begin{tabular}[c]{@{}l@{}}EMO computed\\ values\end{tabular} \cite{hou_line_2018} \\ \hline
$B$  & 0.3324680(42) & 0.33246160 & 0.33248096 \\
$D\times10^7$ \tnote{c} & 4.419 & 4.419  & 4.418  \\
$\gamma$   & 0.001202(70) & 0.0012380  & \\
$\gamma_D\times10^8$ & 1.668 (fixed) & & \\ \hline
 \multicolumn{4}{c}{A$^2\Pi$ $(\nu=0)$} \\ \hline
 Parameters \tnote{a} & This work & \begin{tabular}[c]{@{}l@{}}Mass-scaled $^{40}$CaF\\ Dunham coefficients\end{tabular}  \\ \hline
 $A$ & 72.6156(11)  & &     \\
 $B$ & 0.337231(24) & 0.033721835  \\
 $D\times10^7$ \tnote{c} & 4.474 & 4.474   \\
 $A_D\times10^4$ & $-$1.07(58) & & \\
 $p + 2q$  & $-$0.0431(21) & &\\
 $q$ & $-$0.00035(36) & &  \\
 $T_0$ & 16528.73539(56) & \\ \hline\hline
\end{tabular}
\begin{tablenotes}\small
    \item [a] All units are in wavenumbers (cm$^{-1}$). The numbers in parentheses represent 2$\sigma$ error estimate in the last quoted decimal point.
    \item [b] The hyperfine and nuclear spin-rotation parameters were held fixed to those found in Ref.~\citenum{anderson_millimeter-wave_1994}: $b_F$ = 0.0037192 cm$^{-1}$, $c$ = 0.001108 cm$^{-1}$, $C_I$ = 9.673$\times$10$^{-7}$ cm$^{-1}$.
    \item [c] Centrifugal distortion constants $D$’’ and $D$’ kept fixed to the mass-scaled $^{40}$CaF Dunham coefficient values. 
\end{tablenotes}
\end{threeparttable}
\end{table}

The root mean square of the fit residuals was calculated as 0.000754 cm$^{-1}$ (23~MHz), which is commensurate with the estimated measurement uncertainty. The complete observed line list for all three isotopologues is included in the Supplementary Information.

Our experimental data provide a slight improvement in the \ce{^40CaF} $A$ state constants (Table \ref{tbl:40caf}) due to the narrow linewidths of the observed lines, compared to the most recent work by Lavy et al \cite{lavy_line_2025}. Additionally, despite the small number of observed lines for \ce{^42CaF} and \ce{^44CaF}, our experimentally determined parameters are in excellent agreement with mass-scaled Dunham coefficients and the computed $X$ state constants by Hou and Bernath \citep{hou_line_2018}.

\begin{figure}[ht]
    \begin{subfigure}[b]{0.5\textwidth}
    \centering
        \includegraphics[width=\textwidth]{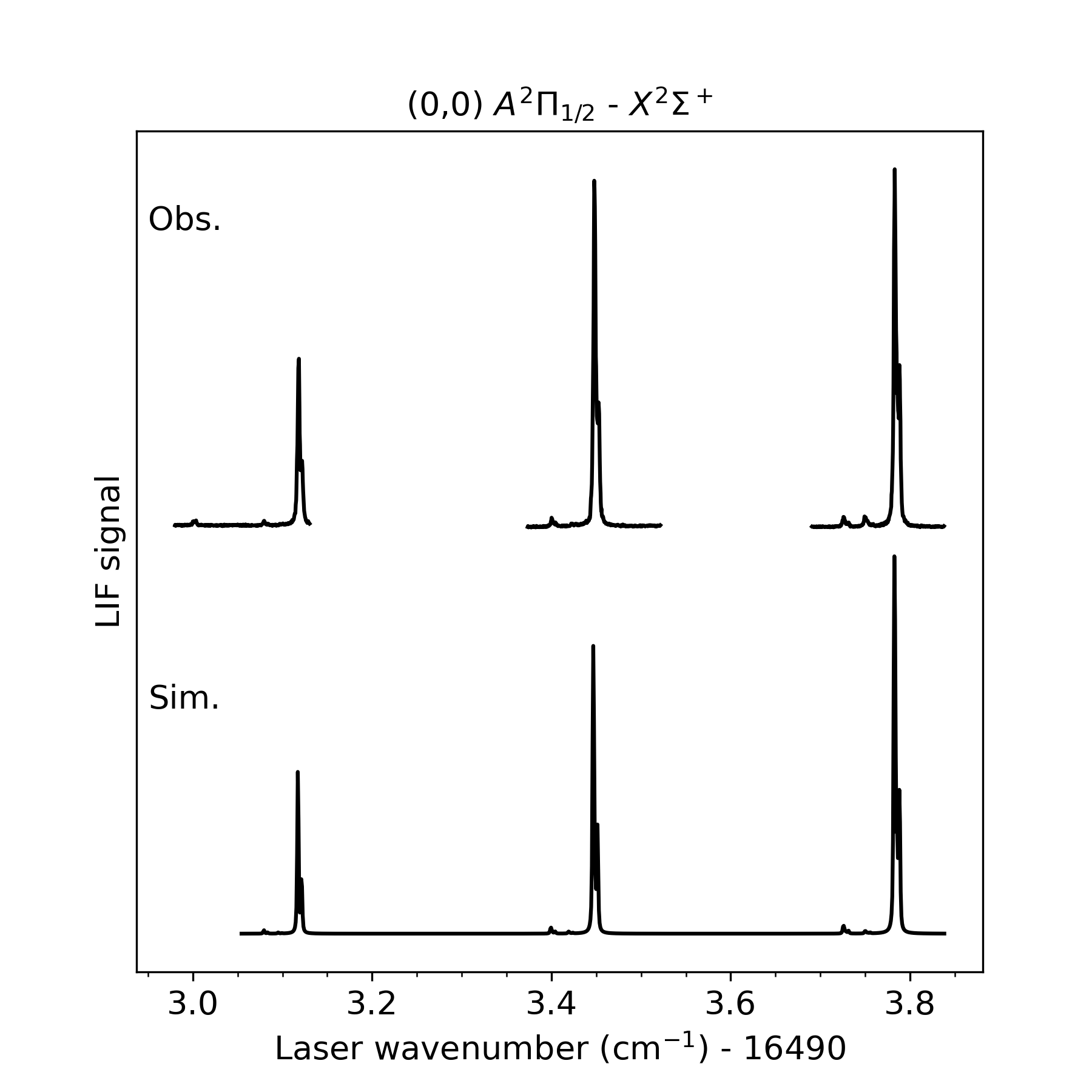}
        \caption{}
    \end{subfigure}
    \begin{subfigure}[b]{0.4\textwidth}
        \centering
        \includegraphics[width=0.83\textwidth]{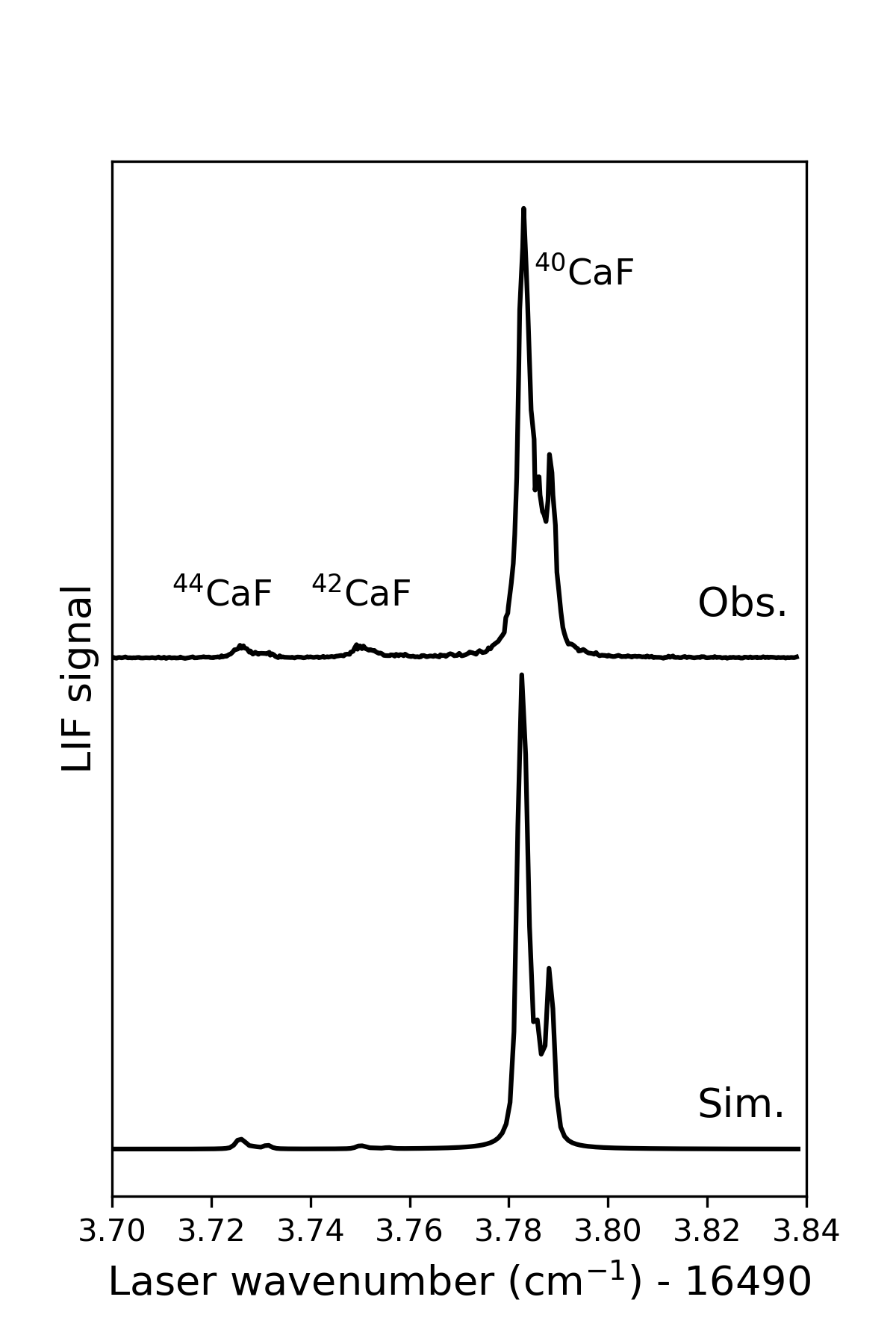}
        \caption{}
    \end{subfigure}
    \caption{(a) The observed and simulated spectra in the regions of the low-$J$ $Q_1$ and $^QR_{12}$ branch features of (0,0) $A^2\Pi$ -- $X^2\Sigma^+$ band system. (b) Enlarged spectrum of the $Q_1(2)$ and $^QR_{12}(2)$ sets of lines, showing isotopologues \ce{^40CaF}, \ce{^42CaF}, and \ce{^44CaF}.}

    \label{fig:lowJ}
\end{figure}

The observed spectra in the regions of the low-$J$ $Q_1$ and $^QR_{12}$ branches of the (0,0) $A^2\Pi$~--~$X^2\Sigma^+$ band system are shown in Figure \ref{fig:lowJ}a. The well-resolved \ce{^40CaF} lines exhibit splittings due to the \ce{^19F} magnetic hyperfine interaction in the ground state, as shown in Figure \ref{fig:lowJ}b. Furthermore, the simulated spectra obtained using the optimized spectroscopic parameters presented in Tables \ref{tbl:40caf}, \ref{tbl:42caf}, and \ref{tbl:44caf}, convoluted with a Voigt function of 40 MHz FWHM, are shown in this figure. 

There were no attempts to measure accurate absolute line intensities or to investigate their isotopologue dependence, as the goal was to observe transitions of less-abundance isotopologues and to provide experimentally determined spectroscopic constants.

\subsection{Isotope abundance ratios}
In principle, the isotope abundance ratios of \ce{^40Ca} and \ce{^44Ca}, $r$(\ce{^44Ca}/\ce{^40Ca}), can be obtained by integrating the spectral line intensities. Ideally, alternating measurements would be performed on an unknown and a reference sample to compensate for systematic isotope fractionation, e.g. in the laser ablation process. This work primarily aimed to determine molecular constants, and the isotope ratio measurements were secondary. Although we did not control possible sources of systematic errors, we can explore the extent to which the isotope ratios can be determined from the obtained data.

Preliminary estimates of the isotope abundance ratio were obtained by fitting the same sets of lines for both \ce{^40CaF} and \ce{^44CaF} isotopologues to a Voigt profile using PGOPHER line positions, floating the width and height parameters. The areas under each curve were integrated and the relative isotope abundance was estimated by taking the ratio of the integrated areas of \ce{^40CaF} and \ce{^44CaF}. 
To minimize the influence of scan-to-scan variability, only pairs of \ce{^40CaF} and \ce{^44CaF} lines that appeared in the same spectral scan were used to determine the isotope ratio. 

\begin{figure}[ht]
    \begin{subfigure}[b]{0.5\textwidth}
    \centering
        \includegraphics[width=\textwidth]{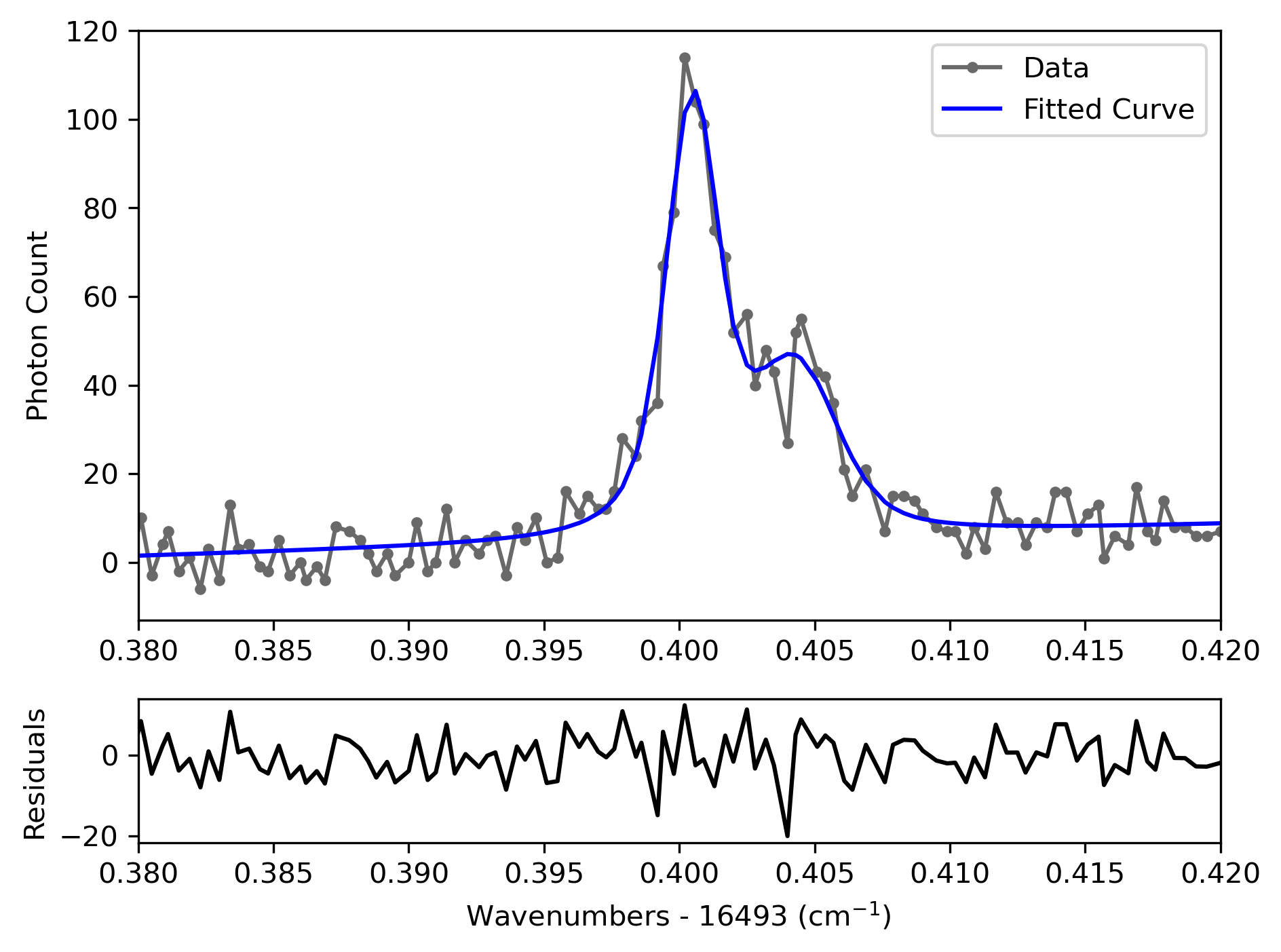}
        \caption{}
    \end{subfigure}
    \begin{subfigure}[b]{0.5\textwidth}
        \centering
        \includegraphics[width=\textwidth]{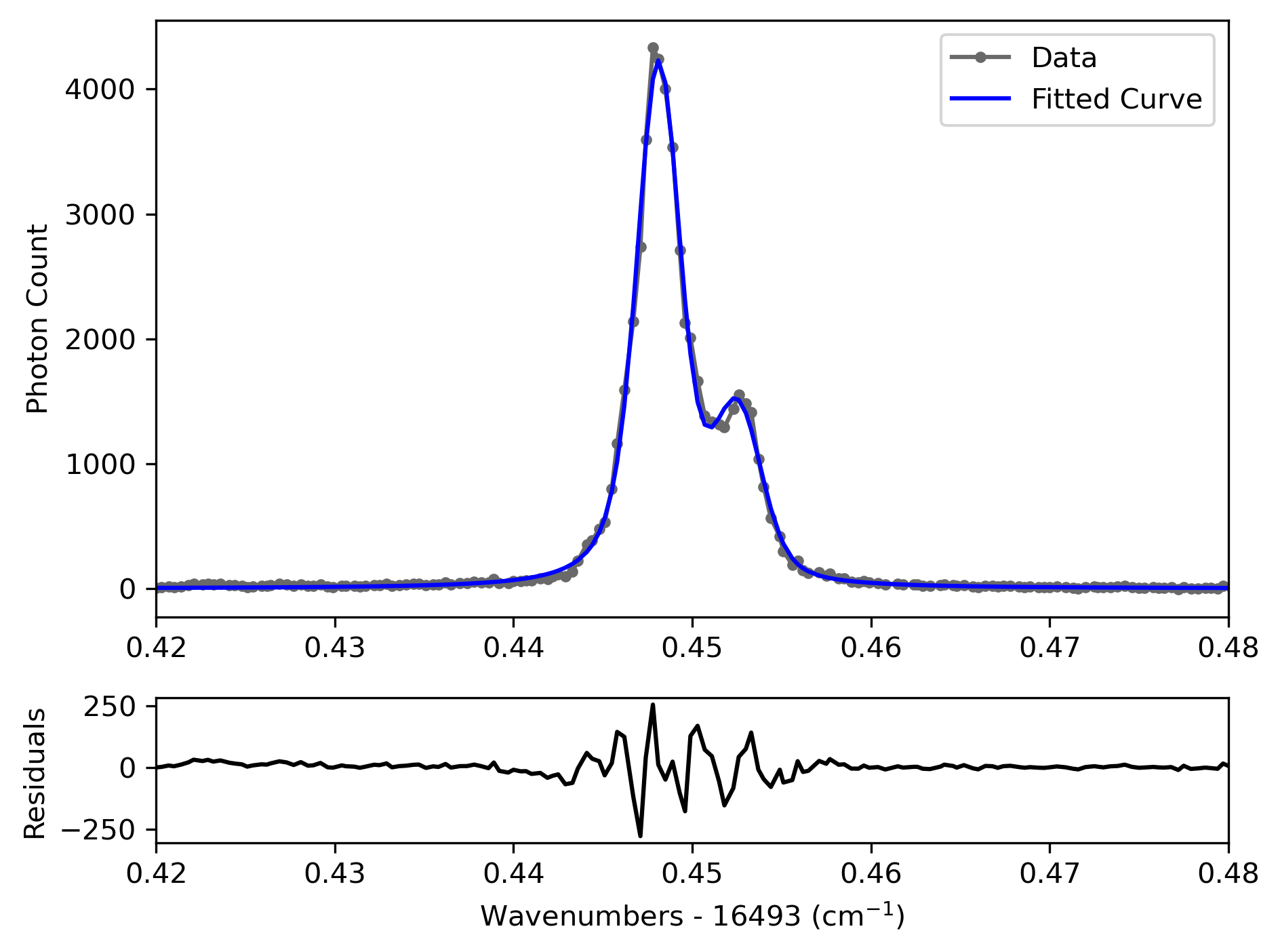}
        \caption{}
    \end{subfigure}
    \caption{(a) The observed $Q_{1}(1)$ and $^QR_{12}(1)$ lines of \ce{^44CaF} and the fitted curve. (b) The observed $Q_{1}(1)$ and $^QR_{12}(1)$ lines of \ce{^40CaF} and the fitted curve.}

    \label{fig:fit}
\end{figure}

Figure \ref{fig:fit} shows lines $Q_{1}(1)$ and $^QR_{12}(1)$ of \ce{40CaF} and \ce{^44CaF}. The experimental isotope ratio determined from these lines is $r$(\ce{^44Ca}/\ce{^40Ca}) = 0.0213(2). This procedure was repeated for lines $Q_{1}(0)$, $Q_{1}(2)$ and $^QR_{12}(2)$, and $R_{2}(1)$ and $^RQ_{21}(1)$ (see Table \ref{tbl:ratios} and Figures S1-S3 in the SI). The weighted mean of all four measurements is 0.0222(6).

\begin{table}[ht]
\centering
\caption[]{Measured isotope ratios of select lines and their statistical summary}\label{tbl:ratios}
\begin{threeparttable}[b]
\centering
\begin{tabular}{llll}\hline 
Transitions & \begin{tabular}[c]{@{}l@{}}Measured isotope ratio\\ $r$(\ce{^44Ca}/\ce{^40Ca}) $\pm$ 1$\sigma$\end{tabular} & $\delta^{44/40}$Ca (\textperthousand)\\ \hline
$Q_1(0)$                   &   0.02016 $\pm$ 0.00023 &  $-$62.32 $\pm$ 10.63\\ 
$Q_1(1)$ and $^QR_{12}(1)$ &  0.02134 $\pm$ 0.00022  &  $-$7.44 $\pm$ 10.36 \\ 
$Q_1(2)$ and $^QR_{12}(2)$ & 0.02530 $\pm$ 0.00029   &  +176.7 $\pm$ 13.41\\
$R_2(1)$ and $^RQ_{21}(1)$ & 0.02583 $\pm$ 0.00042&  +201.4 $\pm$ 19.42\\ \hline
\textbf{Weighted mean} & \textbf{0.02224 $\pm$ 0.0006} & \textbf{+34.42 $\pm$ 6.16}\\
\hline\hline
\end{tabular}
\end{threeparttable}
\end{table}

The known isotope ratio of the NIST SRM 915a calcium carbonate, given in Ref. \citenum{coplen_compilation_2002}, is 0.02152(4). Our weighted mean value agrees to within two standard errors of the NIST reference sample isotope ratio.

Calcium isotopic composition is commonly expressed in delta notation, as a difference in $r$(\ce{^44Ca}/\ce{^40Ca}) of a sample and $r$(\ce{^44Ca}/\ce{^40Ca}) of a standard reference (NIST SRM 915a in this case), in parts per thousand:
\[
\delta^{44/40}\ce{Ca} = \Big(\frac{r(\ce{^44Ca}/\ce{^40Ca})_{sample}}{r(\ce{^44Ca}/\ce{^40Ca})_{reference}}-1\Big)\times 1000
\]

The delta notation enables direct comparison of samples analyzed by different techniques. The delta values reported in Table~\ref{tbl:ratios} have high precision for individual peaks. However, the overall accuracy is limited by systematic errors, particularly those affecting the \ce{^44CaF} signal-to-noise ratio (SNR). While improved SNR would improve accuracy, our current limitations appear to arise primarily from systematic fluctuations.

Sensitivity can be improved through increased signal averaging, optimization of beam conditions, and refinement of the optical layout. A major source of noise is shot-to-shot variation in the ablation, which could be mitigated by normalization procedures.

Several potential sources of systematic error have not yet been fully addressed:
\begin{itemize}
\item Deviations in the sample's isotopic composition from the certified NIST SRM 915a
\item Isotope fractionation during ablation or timing effects in the plume evolution
\item Mass-dependent hydrodynamic effects, including changes in flow conditions across different measurement days
\item Laser power fluctuations during tuning, which could introduce intensity biases between \ce{^44CaF} and \ce{^40CaF} peaks. This can be corrected by normalizing to the laser pulse energy.
\end{itemize}
Alternating measurements between the sample and a known reference target could compensate for many of these systematic effects.

The primary sources of biases from spectroscopy are likely to be small. Differences in transition intensities due to angular momentum coupling are accounted for in the PGOPHER simulations. While we recoded the transitions from $v=0$ levels, small differences in vibrational energies could slightly affect the partition functions. At typical source temperatures ($\sim$100 K), these effects are negligible, though deviations from a Boltzmann distribution could introduce bias.
We have calculated Franck–Condon factors and find they contribute less than 0.1$\%$ deviation. While these were not included in the present analysis, they could be readily incorporated in future work.

The dependence of mass spectrometry techniques on certified reference materials can introduce biases in the determined isotope ratios. Our preliminary results suggest that, in principle, a spectroscopic method could determine calcium isotope ratios from absolute intensity measurements, without the need for reference. Therefore, developing methods that provide robust and unbiased absolute ratios without relying on such standards would be highly valuable for calcium isotope analysis.

\section{Conclusions}

In this study, we experimentally determined the spectroscopic parameters for the $A^2\Pi$~--~$X^2\Sigma^+$ band system of \ce{^40CaF}, \ce{^42CaF}, and \ce{^44CaF}, representing the first experimental investigation of the less-abundant CaF isotopologues. By recording spectra near the natural linewidth limit and with sufficient sensitivity, we were able to compare fine and \ce{^19F} magnetic hyperfine parameters across the three isotopologues. Additionally, we provided a preliminary estimate of the \ce{^44Ca}/\ce{^40Ca} isotopic ratio from the spectral intensities of multiple lines. These findings contribute to the growing body of work on calcium monofluoride spectroscopy, extending it beyond the main isotopologue, \ce{^40CaF}.

Future work could refine these results by measuring isotopically enriched samples to enhance the accuracy of spectroscopic constants for the less abundant isotopologues. Enriched samples can also provide data for other isotopologues not studied here, allowing for a broader investigation of isotopologue dependent effects, such as line intensities. The isotope abundance ratios can be improved by better controlling systematic errors within the experiment, integrating additional sets of spectral lines, and investigating potential mechanisms of isotope fractionation within the molecular beam. Finally, precision measurements of isotope shifts can reveal details about nuclear structure and interaction, including shape parameters and possible new fundamental forces \cite{berengut_probing_2018,hur_evidence_2022,athanasakis-kaklamanakis_king-plot_2023}.

\section{Acknowledgement}

We acknowledge support from the Caltech Space-Health Innovation Fund. T.B. is grateful for support from the Natural Sciences and Engineering Research Council of Canada Postgraduate Scholarship -- Doctoral Program (PGSD3-568209-2022).

\bibliographystyle{elsarticle-num}
\bibliography{Calcium_isotopes}

\end{document}


\maketitle

\listoftables
\listoffigures

\small
\begin{longtable}{llllll|llllll}
\caption{The observed and calculated linelist (cm$^{-1}$) for (0,0) $A^2\Pi_{1/2}$ -- $X^2\Sigma^+$ band system of $^{40}$CaF.}\label{tbl:1SI_40caf} \\ 
\hline
Branch\footnotemark[1] & $J"$ & $F"$ & $F'$ & Obs.\footnotemark[2] & Obs. - Calc.\footnotemark[2] & Branch & $J"$ & $F"$ & $F'$ & Obs. & Obs. - Calc. \\ 
\hline
\endfirsthead

\hline
Branch & $J"$ & $F"$ & $F'$ & Obs. & Obs. - Calc. & Branch & $J"$ & $F"$ & $F'$ & Obs. & Obs. - Calc. \\ 
\hline
\endhead

\hline
\endfoot

$O_{1}(6)$	&	6.5	&	6	&	5	&	486.93907	&	-0.00102	&  $^PQ_{12}(28)$	&	27.5	&	27	&	27	&	486.44744	&	-0.00152	\\
    $O_{1}(7)$	&	7.5	&	7	&	6	&	485.9274	&	0.00103		&                 &	27.5	&	28	&	28	&	486.44959	&	-0.00128	\\
    $O_{1}(8)$	&	8.5	&	8	&	7	&	484.9184	&	-0.00105	& $^PQ_{12}(29)$	&	28.5	&	28	&	28	&	486.3182	&	-0.00109	\\
    $O_{1}(9)$	&	9.5	&	9	&	8	&	483.91958	&	0.00021		&                   &	28.5	&	29	&	29	&	486.3201	&	-0.0011	\\
    $^OP_{12}(6)$	&	5.5	&	5	&	4	&	486.94649	&	-0.00067	& $^PQ_{12}(30)$	&	29.5	&	29	&	29	&	486.19646	&	0.00002	\\
    	&	5.5	&	6	&	5	&	486.94834	&	-0.001		&               &	29.5	&	30	&	30	&	486.198	&	-0.00035	\\
    $^OP_{12}(7)$	&	6.5	&	6	&	5	&	485.9354	&	0.0007	&  $^PQ_{12}(31)$	&	30.5	&	30	&	30	&	486.08152	&	0.00108	\\
    	&	6.5	&	7	&	6	&	485.93669	&	-0.00016		&                 &	30.5	&	31	&	31	&	486.0828	&	0.00046	\\
    $^OP_{12}(8)$	&	7.5	&	7	&	6	&	484.92785	&	-0.00121	 &$^PQ_{12}(32)$	&	31.5	&	31	&	31	&	485.97182	&	0.00051	\\
    	&	7.5	&	8	&	7	&	484.93028	&	-0.00089		&                &	31.5	&	32	&	32	&	485.97341	&	0.00021	\\
    $^OP_{12}(9)$	&	8.5	&	8	&	7	&	483.9306	&	0.00033	&   $Q_{1}(0)$	&	0.5	&	1	&	1	&	493.11798	&	0.00071	\\
    	&	8.5	&	9	&	8	&	483.93244	&	0.00009		&                    &	0.5	&	1	&	0	&	493.11798	&	0.00071	\\
    $P_{1}(2)$	&	2.5	&	3	&	2	&	492.16521	&	-0.00088		&                        &	0.5	&	0	&	1	&	493.12153	&	0.00018	\\
    $P_{1}(24)$	&	24.5	&	25	&	24	&	487.0035	&	0.00044	&          $Q_{1}(1)$	&	1.5	&	2	&	2	&	493.44796	&	0.00145	\\
    	&	24.5	&	24	&	23	&	487.0048	&	-0.00005		&                         &	1.5	&	1	&	1	&	493.44865	&	0.00127	\\
    $P_{1}(25)$	&	25.5	&	26	&	25	&	486.8454	&	0.00047	 & $Q_{1}(2)$	&	2.5	&	3	&	3	&	493.78302	&	0.00066	\\
    	&	25.5	&	25	&	24	&	486.8464	&	-0.00032	 & $Q_{1}(6)$	&	6.5	&	7	&	7	&	495.19347	&	0.00127	\\
    $P_{1}(26)$	&	26.5	&	27	&	26	&	486.69406	&	0.00052		&                         &	6.5	&	6	&	6	&	495.1945	&	0.00077	\\
    	&	26.5	&	26	&	25	&	486.69549	&	0.00014	&         $Q_{1}(8)$	&	8.5	&	9	&	9	&	495.93721	&	0.00008	\\
    $P_{1}(27)$	&	27.5	&	28	&	27	&	486.54887	&	-0.00006		&                         &	8.5	&	8	&	8	&	495.9383	&	-0.00044	\\
	&	27.5	&	27	&	26	&	486.5501	&	-0.00063	&         $Q_{1}(9)$	&	9.5	&	10	&	10	&	496.31911	&	-0.00053	\\
    $P_{1}(28)$	&	28.5	&	29	&	28	&	486.41055	&	-0.00055		&                    &	9.5	&	9	&	9	&	496.31993	&	-0.00135	\\
    	&	28.5	&	28	&	27	&	486.4121	&	-0.00081	& $Q_{12}(1)$	&	0.5	&	0	&	1	&	492.4796	&	0.00054	\\
    $P_{1}(29)$	&	29.5	&	30	&	29	&	486.28007	&	0		&                      &	0.5	&	1	&	1	&	492.48265	&	0.00096	\\
    	&	29.5	&	29	&	28	&	486.28085	&	-0.00103	 &     $^QR_{12}(1)$	&	0.5	&	1	&	2	&	493.45276	&	0.00132	\\
    $P_{1}(30)$	&	30.5	&	31	&	30	&	486.15681	&	0.00095	&      $^QR_{12}(2)$	&	1.5	&	1	&	2	&	493.78586	&	0.00008	\\
    	&	30.5	&	30	&	29	&	486.15854	&	0.00086		&                       &	1.5	&	2	&	3	&	493.78842	&	0.00009	\\
    $P_{1}(31)$	&	31.5	&	32	&	31	&	486.03964	&	0.00114	&  $^QR_{12}(6)$	&	5.5	&	5	&	6	&	495.2022	&	0.00141	\\
    	&	31.5	&	31	&	30	&	486.04117	&	0.00085		&                      &	5.5	&	6	&	7	&	495.20432	&	0.00134	\\ 
    $P_{1}(32)$	&	32.5	&	33	&	32	&	485.92891	&	0.00091	&      $^QR_{12}(8)$	&	7.5	&	7	&	8	&	495.94851	&	0.00016	\\
    $^QP_{1}(1)$	&	1.5	&	2	&	1	&	492.4773	&	0.00054		&                      &	7.5	&	8	&	9	&	495.94988	&	-0.00058	\\
	&	1.5	&	1	&	0	&	492.47838	&	0.00075	&  $^QR_{12}(9)$	&	8.5	&	8	&	9	&	496.33173	&	-0.00045	\\
	&	1.5	&	1	&	1	&	492.4785	&	0.00087		&                        &	8.5	&	9	&	10	&	496.333	&	-0.00126	\\
    $^PQ_{12}(2)$	&	1.5	&	1	&	1	&	492.16852	&	-0.00099	&        $R_{1}(1)$	&	1.5	&	2	&	3	&	495.2887	&	0.00058	\\
    	&	1.5	&	2	&	2	&	492.17057	&	-0.00148	&      $R_{1}(2)$	&	2.5	&	3	&	4	&	496.3593	&	-0.0012	\\
    $^PQ_{12}(24)$	&	23.5	&	23	&	23	&	487.0355	&	-0.00003		&                  &	2.5	&	2	&	3	&	496.36016	&	-0.00147	\\
	&	23.5	&	24	&	24	&	487.03742	&	-0.00004	    &$^RS_{12}(1)$	&	0.5	&	1	&	2	&	495.29331	&	0.00025	\\
    $^PQ_{12}(26)$	&	25.5	&	25	&	25	&	486.729	&	0.00029	&      $^RS_{12}(2)$	&	1.5	&	2	&	3	&	496.36538	&	-0.00109	\\
    	&	25.5	&	26	&	26	&	486.73059	&	-0.00003		\\										
    $^PQ_{12}(27)$	&	26.5	&	26	&	26	&	486.5854	&	-0.00004		\\										
    	&	26.5	&	27	&	27	&	486.5868	&	-0.00056		\\

\end{longtable}

\small
\begin{longtable}{llllll|llllll}
\caption{The observed and calculated linelist (cm$^{-1}$) for (0,0) $A^2\Pi_{3/2}$ -- $X^2\Sigma^+$ band system of $^{40}$CaF.}\label{tbl:1SI_40caf} \\ 
\hline
Branch\footnotemark[1] & $J"$ & $F"$ & $F'$ & Obs.\footnotemark[2] & Obs. - Calc.\footnotemark[2] & Branch & $J"$ & $F"$ & $F'$ & Obs. & Obs. - Calc. \\ 
\hline
\endfirsthead

\hline
Branch & $J"$ & $F"$ & $F'$ & Obs. & Obs. - Calc. & Branch & $J"$ & $F"$ & $F'$ & Obs. & Obs. - Calc. \\ 
\hline
\endhead

\hline
\endfoot	
	$^PO_{21}(6)$	&	6.5	&	6	&	5	&	559.05102	&	-0.00024&	 $^RQ_{21}(3)$	&	3.5	&	4	&	4	&	566.18610	&	0.00104 \\
    $^PO_{21}(7)$	&	7.5	&	7	&	6	&	558.09433	&	-0.00040&	                &	3.5	&	3	&	3	&	566.18700	&	0.00065 \\
    $^PO_{21}(9)$	&	9.5	&	9	&	8	&	556.22024	&	-0.00042&                     &	3.5	&	3	&	4	&	566.18700	&	0.00065 \\
    $P_{2}(6)$	     &	5.5	&	5	&	4	&	559.05840	&	0.00007	&	$^RQ_{21}(26)$	&	26.5&	27	&	27	&	578.59703	&	0.00100	\\
            	    &	5.5	&	6	&	5	&	559.06056	&	0.00005	&	             &	26.5	&	26	&	26	&	578.59818	&	0.00034	\\
    $P_{2}(7)$	     &	6.5	&	6	&	5	&	558.10309	&	0.00003	&	$^RQ_{21}(32)$	&	32.&	33	&	33	&	582.90902	&	-0.00010	\\
        	       &	6.5	&	7	&	6	&	558.10472	&	-0.00049&	              &	32.5	&	32	&	32	&	582.91031	&	-0.00063	\\
    $P_{2}(9)$	    &	8.5	&	8	&	7	&	556.23187	&	0.00031	&	   $R_{2}(1)$	&	0.5	&	0	&	1	&	565.42602	&	0.00009	\\
        	       &	8.5	&	9	&	8	&	556.23321	&	-0.00043&	               &	0.5	&	1	&	1	&	565.42853	&	-0.00003	\\
    $^QP_{21}(2)$	&	2.5	&	3	&	2	&	564.05301	&	-0.00005&	                &	0.5	&	1	&	2	&	565.42853	&	-0.00003	\\
    	           &	2.5	&	2	&	2	&	564.05301	&	-0.00005&	  $R_{2}(2)$	&	1.5	&	1	&	2	&	565.80130	&	0.00000	\\
    $^QP_{21}(3)$	&	3.5	&	4	&	3	&	563.74206	&	-0.00038&	                &	1.5	&	2	&	3	&	565.80365	&	-0.00019	\\
        	       &	3.5	&	3	&	3	&	563.74282	&	-0.00091&	                  &	1.5	&	2	&	2	&	565.80365	&	-0.00019	\\	
        	       &	3.5	&	3	&	2	&	563.74282	&	-0.00091&	  $R_{2}(3)$	&	2.5	&	2	&	3	&	566.19060	&	0.00085	\\
    $^QP_{21}(4)$	&	4.5	&	5	&	4	&	563.44570	&	0.00084	&	                &	2.5	&	3	&	4	&	566.19258	&	0.00042	\\	
        	       &	4.5	&	4	&	3	&	563.44686	&	0.00061	&	                  &	2.5	&	3	&	3	&	566.19258	&	0.00042	\\
    	           &	4.5	&	4	&	4	&	563.44686	&	0.00061	&	 $R_{2}(26)$&	25.5	&	25	&	26	&	578.63206	&	0.00086	\\
    $^QP_{21}(5)$	&	5.5	&	6	&	5	&	563.15957	&	-0.00075&	            &	25.5	&	26	&	27	&	578.63355	&	0.00044	\\	
        	       &	5.5	&	5	&	4	&	563.16059	&	-0.00119&	$R_{2}(32)$	&	31.5	&	31	&	32	&	582.95190	&	-0.00053	\\
        	       &	5.5	&	5	&	5	&	563.16059	&	-0.00119&		         &	31.5	&	32	&	33	&	582.95340	&	-0.00093	\\
    $Q_{2}(2)$	 &	1.5	&	1	&	1	&	    564.05621	&	-0.00027	&	$^SR_{21}(0)$	&	0.5	&	1	&	2 & 566.10957	&	0.00033	\\	
        	       &	1.5	&	2	&	2	&	564.05830	&	-0.00073&	                  &	0.5	&	0	&	1	&	566.11399	&	0.00066	\\
        	       &	1.5	&	2	&	1	&	564.05830	&	-0.00073&	$^SR_{21}(11)$	&	11.5	&	12	&	13& 578.47413	&	0.00094	\\
    $Q_{2}(3)$	      &	2.5	&	2	&	2	&	563.74716	&	0.00002	&	              &	11.5	&	11	&	12	&	578.47544	&	0.00057	\\	
        	       &	2.5	&	3	&	3	&	563.74875	&	-0.00079&	 $^SR_{21}(12)$	&12.5	&	13	&	14	&	579.67621	&	0.00171	\\
    $Q_{2}(4)$	     &	3.5	&	3	&	3	&	563.45170	&	0.00086	&	              &	12.5	&	12	&	13	&	579.67770	&	0.00150	\\
        	       &	3.5	&	4	&	4	&	563.45385	&	0.00070	&	$^SR_{21}(15)$	&	15.5&	16	&	17	&	583.35453	&	-0.00068	\\
    $Q_{2}(5)$	&	4.5	&	4	&	4	&	    563.16698	&	-0.00062	&$^SR_{21}(15)$	&	15.5	&	15	&16	&	583.35560	&	-0.00134	\\
        	       &	4.5	&	5	&	5	&	563.16876	&	-0.00108	& $S_{2}(11)$	&	10.5	&	11	&	12&	578.49102	&	0.00061	\\
    $^RQ_{21}(1)$	&	1.5	&	2	&	2	&	565.42409	&	0.00047  & $S_{2}(12)$	&	11.5	&	12	&	13	&	579.69437	&	0.00135	\\
        	       &	1.5	&	1	&	2	&	565.42412	&	-0.00037	& $S_{2}(15)$	&	14.5	&	15	&	16&	583.37712	&	-0.00053	\\
    $^RQ_{21}(2)$	&	2.5	&	3	&	3	&	565.79796	&	0.00008 \\
        	       &	2.5	&	2	&	3	&	565.79860	&	-0.00040 \\
        	       &	2.5	&	2	&	2	&	565.79860	&	-0.00040 \\

\end{longtable}

\begin{table}[h] 
\caption{The observed and calculated linelist (cm$^{-1}$) for (0,0) $A^2\Pi_{1/2}$ -- $X^2\Sigma^+$ band system of $^{42}$CaF.}\label{tbl:3SI_42caf}
\small
\begin{threeparttable}
\centering
\begin{tabular}{llllll|llllll}\hline
Branch\footnotemark[1] & $J"$ & $F"$ & $F'$ & Obs.\footnotemark[2] & Obs. - Calc.\footnotemark[2] & Branch & $J"$ & $F"$ & $F'$ & Obs. & Obs. - Calc. \\	\hline
	$^OP_{12}(6)$	&	6.5	&	8	&	7	&	485.03485	&	-0.00035& $Q_{1}(6)$	&	6.5	&	7	&	7	&	495.14379	&	0.00044	\\
    $^OP_{12}(7)$	&	7.5	&	9	&	8	&	484.05365	&	0.00105	&         	&	6.5	&	6	&	6	&	495.14379	&	-0.00107 \\	
    $^OP_{12}(8)$	&	8.5	&	10	&	9	&	483.07592	&	-0.00117&       $^RQ_{1}(1)$	&	0.5	&	1	&	0	&	493.09698	&	-0.00040\\
    $^OP_{12}(8)$	&	8.5	&	9	&	8	&	483.07592	&	0.00090	&    	&	0.5	&	0	&	1	&	493.10059	&	-0.00087\\
    $Q_{1}(1)$	    &	1.5	&	2	&	2	&	493.42280	&	0.00147	& $^QR_{12}(1)$	&	1.5	&	1	&	2	&	493.42770	&	0.00148	\\
        	&	1.5	&	1	&	1	&	493.42280	&	0.00147	& $^QR_{12}(6)$	&	6.5	&	6	&	7	&	495.15280	&	-0.00103\\
    $Q_{1}(2)$	&	2.5	&	3	&	3	&	493.75021	&	-0.00186	& &	6.5	&	5	&	6	&	495.15280	&	-0.00103\\ 
          
\end{tabular}
\end{threeparttable}
\end{table}

\begin{table}[h] 
\caption{The observed and calculated linelist (cm$^{-1}$) for (0,0) $A^2\Pi_{3/2}$ -- $X^2\Sigma^+$ band system of $^{42}$CaF.}\label{tbl:4SI_42caf}
\small
\begin{threeparttable}
\centering
\begin{tabular}{llllll|llllll}\hline
Branch\footnotemark[1] & $J"$ & $F"$ & $F'$ & Obs.\footnotemark[2] & Obs. - Calc.\footnotemark[2] & Branch & $J"$ & $F"$ & $F'$ & Obs. & Obs. - Calc. \\\hline	
$P_{2}(5)$	&	4.5	&	6	&	5	&	559.13881	&	-0.00080	& $R_{2}(2)$	&	1.5	&	1	&	1	&	565.40865	&	-0.00041	\\
$P_{2}(6)$	&	5.5	&	6	&	5	&	558.19766	&	-0.00062	& 	&	1.5	&	1	&	2	&	565.40865	&	-0.00041	\\
	&	5.5	&	7	&	6	&	558.20172	&	0.00128	& $^SR_{21}(17)$	&	16.5	&	16	&	17	&	583.12412	&	0.00088	\\
$Q_{2}(5)$	&	4.5	&	5	&	5	&	563.18645	&	-0.00137	&	&	16.5	&	15	&	16	&	583.12412	&	0.00088	\\
$^RQ_{21}(3)$	&	2.5	&	3	&	3	&	565.77524	&	0.00149	\\
	&	2.5	&	2	&	3	&	565.77524	&	0.00149 \\
	&	2.5	&	2	&	2	&	565.77524	&	0.00149	\\
     
\end{tabular}
\end{threeparttable}
\end{table}

\small
\begin{longtable}{llllll|llllll}
\caption{The observed and calculated linelist (cm$^{-1}$) for (0,0) $A^2\Pi_{1/2}$ -- $X^2\Sigma^+$ band system of $^{44}$CaF.}\label{tbl:1SI_40caf} \\ 
\hline
Branch\footnotemark[1] & $J"$ & $F"$ & $F'$ & Obs.\footnotemark[2] & Obs. - Calc.\footnotemark[2] & Branch & $J"$ & $F"$ & $F'$ & Obs. & Obs. - Calc. \\ 
\hline
\endfirsthead

\hline
Branch & $J"$ & $F"$ & $F'$ & Obs. & Obs. - Calc. & Branch & $J"$ & $F"$ & $F'$ & Obs. & Obs. - Calc. \\ 
\hline
\endhead

\hline
\endfoot
$^OP_{12}(4)$	&	4.5	&	5	&	4	&	487.08930	&	-0.00138	&$Q_1(8)$  	   &	8.5	&	9	&	9	&	495.82000	&	-0.00023	\\
            	&	4.5	&	6	&	5	&	487.09210	&	-0.00081	&               &	8.5	&	8	&	8	&	495.82190	&	0.00010	\\
$^OP_{12}(5)$	&	5.5	&	6	&	5	&	486.10820	&	-0.00002	&$Q_1(9)$  	    &	9.5	&	10	&	10	&	496.19200	&	-0.00024	\\
            	&	5.5	&	7	&	6	&	486.11060	&	0.00020	&                     &	9.5	&	9	&	9	&	496.19370	&	-0.00014	\\
$^OP_{12}(7)$	&	7.5	&	8	&	7	&	484.16294	&	-0.00039	&$^RQ_{1}(1)$	&	0.5	&	1	&	1	&	493.07907	&	-0.00031	\\
            	&	7.5	&	9	&	8	&	484.16480	&	-0.00065	&              &	0.5	&	1	&	0	&	493.07907	&	-0.00031	\\
$^OP_{12}(8)$	&	8.5	&	9	&	8	&	483.20137	&	0.00040	&                  &	0.5	&	0	&	1	&	493.08290	&	-0.00057	\\
            	&	8.5	&	10	&	9	&	483.20400	&	0.00094	&     $^QR_{12}(2)$	&	2.5	&	2	&	3	&	493.73151	&	0.00018	\\
$Q_1(1)$  	    &	1.5	&	2	&	2	&	493.40034	&	0.00110	&     $^QR_{12}(6)$	&	6.5	&	6	&	7	&	495.10750	&	0.00152	\\
          	    &	1.5	&	1	&	1	&	493.40060	&	0.00056	&     $^QR_{12}(7)$	&	7.5	&	6	&	7	&	495.46400	&	0.00015	\\
          	    &	1.5	&	1	&	2	&	493.40450	&	0.00042	&                     &	7.5	&	7	&	8	&	495.46554	&	-0.00049	\\
$Q_1(2)$  	    &	2.5	&	3	&	3	&	493.72571	&	0.00014	&     $^QR_{12}(8)$	&	8.5	&	7	&	8	&	495.83087	&	0.00037	\\
          	    &	2.5	&	2	&	2	&	493.72620	&	-0.00042&                     &	8.5	&	8	&	9	&	495.83190	&	-0.00074	\\
$Q_1(6)$  	    &	6.5	&	7	&	7	&	495.09724	&	0.00135	&     $^QR_{12}(9)$	&	9.5	&	8	&	9	&	496.20441	&	0.00070	\\
          	    &	6.5	&	6	&	6	&	495.09809	&	0.00071	&                     &	9.5	&	9	&	10	&	496.20560	&	-0.00022	\\
$Q_1(7)$  	    &	7.5	&	8	&	8	&	495.45371	&	-0.00107\\
         	    &	7.5	&	7	&	7	&	495.45546	&	-0.00086\\

\end{longtable}

\small
\begin{longtable}{llllll|llllll}
\caption{The observed and calculated linelist (cm$^{-1}$) for (0,0) $A^2\Pi_{3/2}$ -- $X^2\Sigma^+$ band system of $^{44}$CaF.}\\ 
\hline
Branch\footnotemark[1] & $J"$ & $F"$ & $F'$ & Obs.\footnotemark[2] & Obs. - Calc.\footnotemark[2] & Branch & $J"$ & $F"$ & $F'$ & Obs. & Obs. - Calc. \\ 
\hline
\endfirsthead

\hline
Branch & $J"$ & $F"$ & $F'$ & Obs. & Obs. - Calc. & Branch & $J"$ & $F"$ & $F'$ & Obs. & Obs. - Calc. \\ 
\hline
\endhead

\hline
\endfoot	
$P_{2}(5)$	&	4.5	&	5	&	4	&	559.21394	&	0.00066	&  $Q_{2}(4)$	&	3.5	&	3	&	3	&	563.47750	&	-0.00002\\
        	&	4.5	&	6	&	5	&	559.21570	&	0.00019	&           	&	3.5	&	4	&	4	&	563.47990	&	0.00002	\\
$P_{2}(6)$	&	5.5	&	6	&	5	&	558.28590	&	0.00013	&  $Q_{2}(5)$	&	4.5	&	4	&	4	&	563.20237	&	-0.00006	\\
        	&	5.5	&	7	&	6	&	558.28780	&	-0.00015	&            &	4.5	&	5	&	5	&	563.20366	&	-0.00105	\\
$P_{2}(8)$	&	7.5	&	8	&	7	&	556.46820	&	-0.00032	&$Q_{2}(6)$	&	5.5	&	5	&	5	&	562.94060	&	0.00060	\\
        	&	7.5	&	9	&	8	&	556.47110	&	0.00046	&              	&	5.5	&	6	&	6	&	562.94190	&	-0.00033	\\
$^QP_{21}(3)$&	2.5	&	4	&	3	&	563.76110	&	0.00014	&$^RQ_{21}(2)$	&	1.5	&	1	&	2	&	565.39390	&	0.00017	\\
        	&	2.5	&	3	&	3	&	563.76180	&	-0.00038	&        	&	1.5	&	1	&	1	&	565.39910	&	0.00133	\\
        	&	2.5	&	3	&	2	&	563.76180	&	-0.00038&$^RQ_{21}(3)$	&	2.5	&	3	&	3	&	565.75669	&	0.00048	\\
$^QP_{21}(4)$&	3.5	&	5	&	4	&	563.47275	&	0.00071	&             	&	2.5	&	2	&	3	&	565.75754	&	0.00027	\\
        	&	3.5	&	4	&	4	&	563.47400	&	0.00062	&            	&	2.5	&	2	&	2	&	565.75754	&	0.00027	\\
        	&	3.5	&	4	&	3	&	563.47400	&	0.00062	&$^RQ_{21}(4)$	&	3.5	&	4	&	4	&	566.13230	&	0.00029	\\
$^QP_{21}(5)$&	4.5	&	6	&	5	&	563.19408	&	-0.00168	&        	&	3.5	&	3	&	3	&	566.13303	&	-0.00021	\\
        	&	4.5	&	5	&	4	&	563.19620	&	-0.00098	&        	&	3.5	&	3	&	4	&	566.13303	&	-0.00021	\\
        	&	4.5	&	5	&	5	&	563.19620	&	-0.00098	&$R_{2}(2)$	&	1.5	&	1	&	2	&	565.39390	&	0.00017	\\
$^QP_{21}(6)$&	5.5	&	7	&	6	&	562.93260	&	0.00047	&            	&	1.5	&	1	&	1	&	565.39812	&	0.00035	\\
        	&	5.5	&	6	&	5	&	562.93445	&	0.00083	&$R_{2}(3)$	     &	2.5	&	2	&	3	&	565.76146	&	-0.00051	\\
        	&	5.5	&	6	&	6	&	562.93445	&	0.00083	&                &	2.5	&	2	&	2	&	565.76146	&	-0.00051	\\
$Q_{2}(3)$	&	2.5	&	2	&	2	&	563.76560	&	0.00034	&$R_{2}(4)$	   &	3.5	&	3	&	4	&	566.13790	&	-0.00088	\\
        	&	2.5	&	3	&	3	&	563.76730	&	-0.00043	&        	&	3.5	&	3	&	3	&	566.13790	&	-0.00088	\\

\end{longtable}
\footnotetext[1]{ A $^{\Delta N}\Delta J_{F_i'F_i''}(N'')$ branch notation is used. Note that in case of $F_i'=F_i''$, the redundancy is removed.}
\footnotetext[2]{ Transition wavenumber - 16000 cm$^{-1}$.}

\begin{figure}[h]
    \begin{subfigure}[b]{0.5\textwidth}
    \centering
        \includegraphics[width=\textwidth]{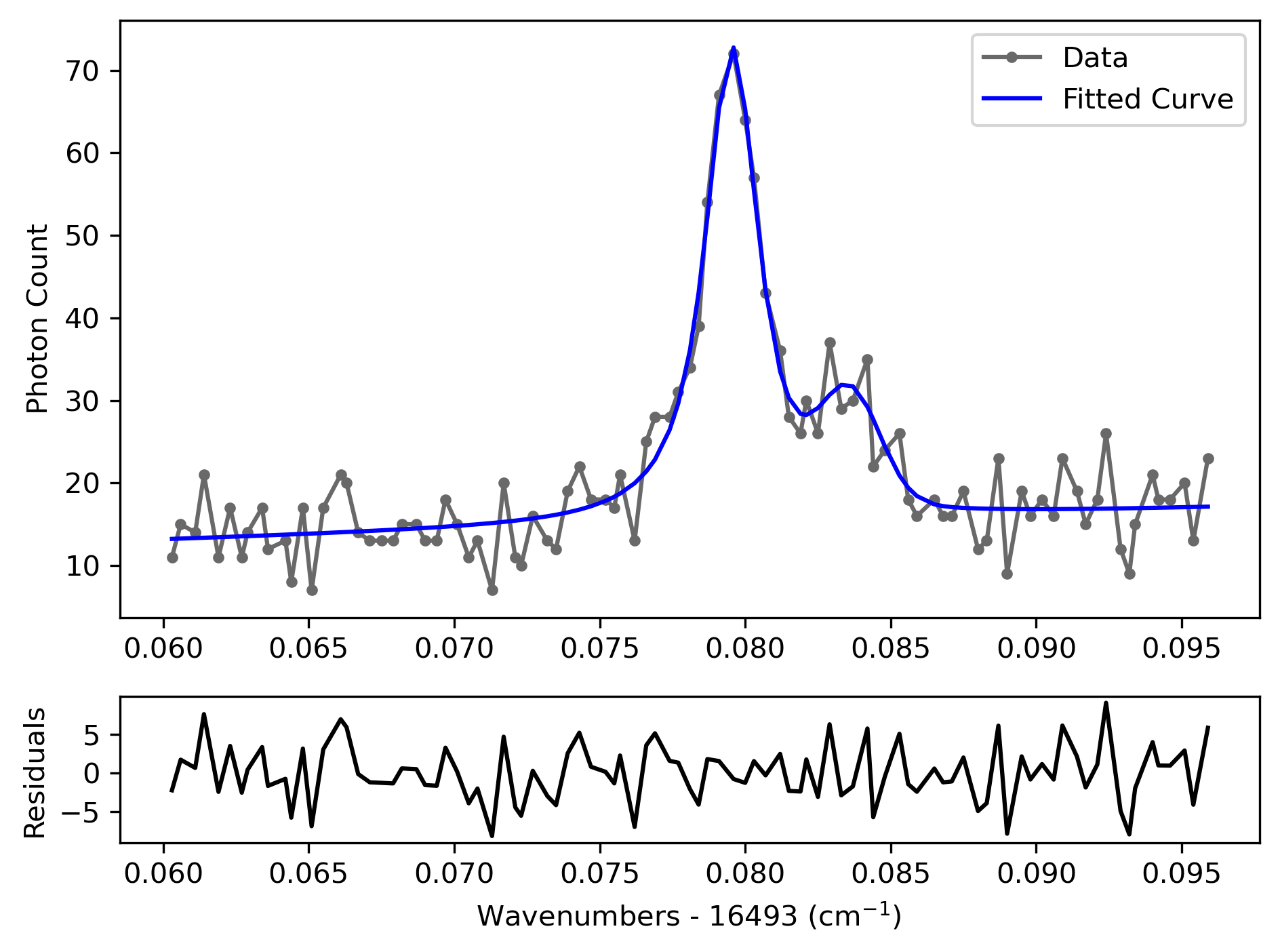}
        \caption{}
    \end{subfigure}
    \begin{subfigure}[b]{0.5\textwidth}
        \centering
        \includegraphics[width=\textwidth]{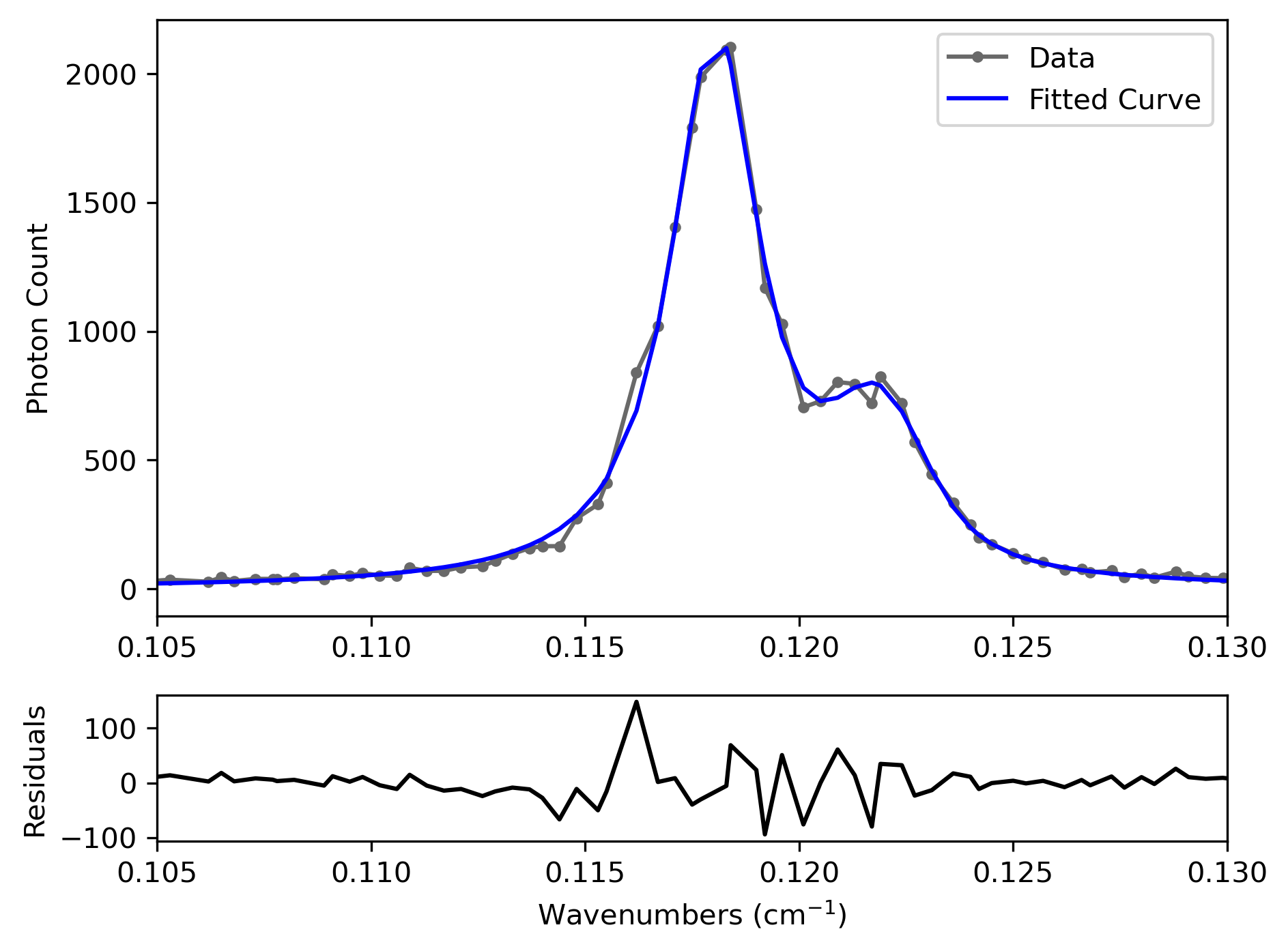}
        \caption{}
    \end{subfigure}
    \caption{(a) The observed $Q_{1}(0)$ lines of \ce{^44CaF} and the fitted curve. (b) The observed $Q_{1}(0)$ lines of \ce{^40CaF} and the fitted curve.}

    \label{fig:fit}
\end{figure}

\begin{figure}[h]
    \begin{subfigure}[b]{0.5\textwidth}
    \centering
        \includegraphics[width=\textwidth]{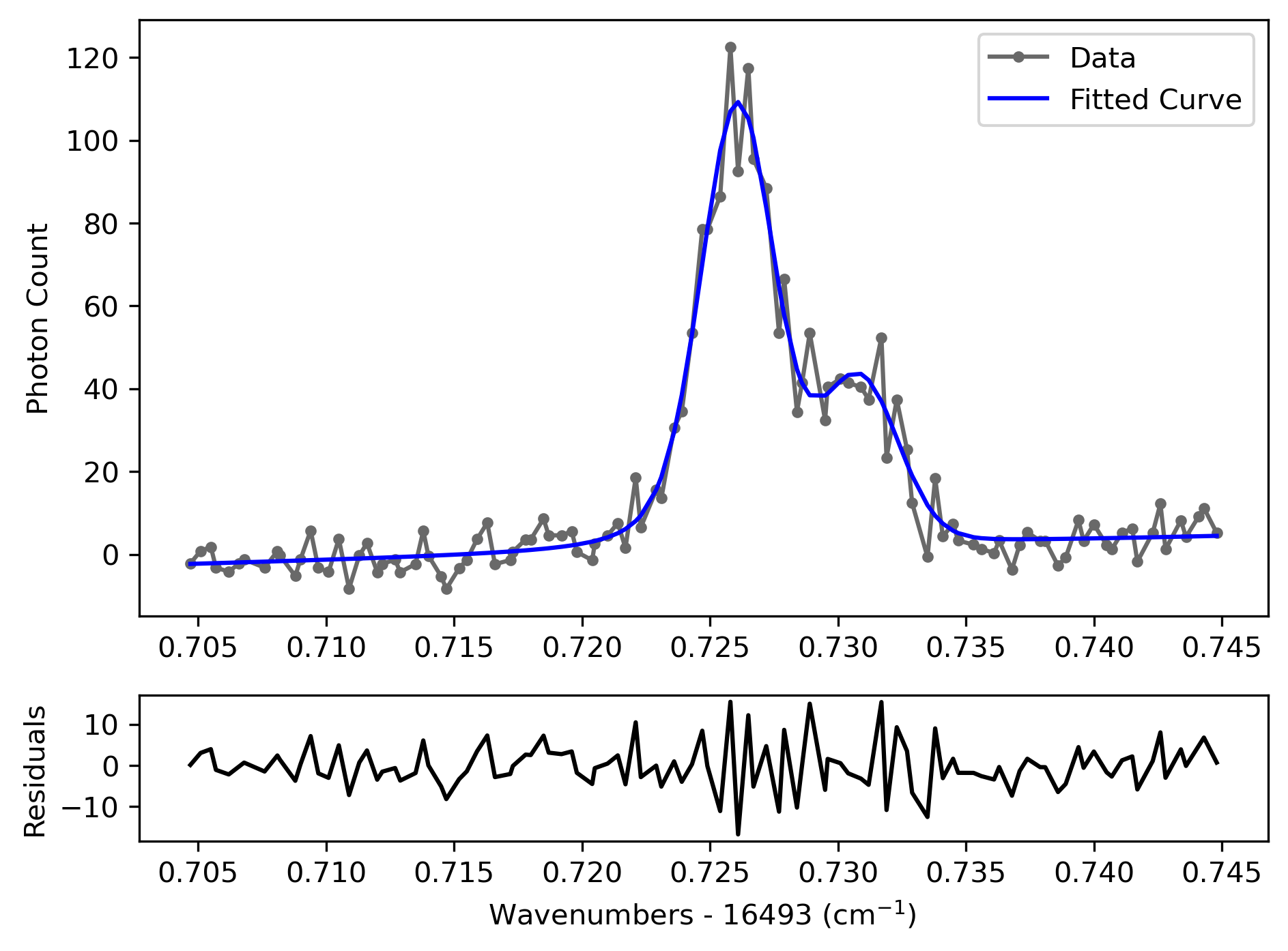}
        \caption{}
    \end{subfigure}
    \begin{subfigure}[b]{0.5\textwidth}
        \centering
        \includegraphics[width=\textwidth]{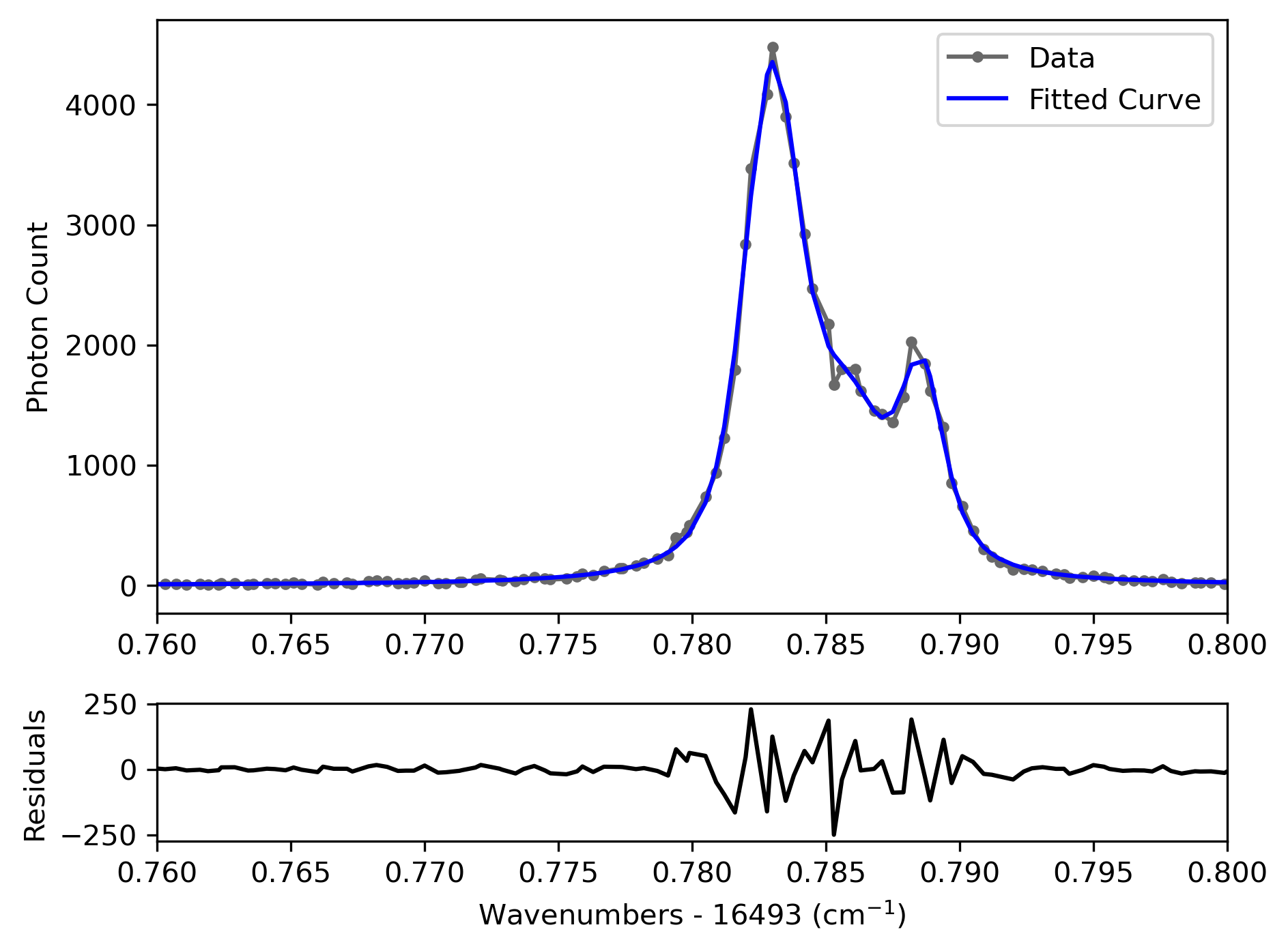}
        \caption{}
    \end{subfigure}
    \caption{(a) The observed $Q_{1}(2)$ and $^QR_{12}(2)$ lines of \ce{^44CaF} and the fitted curve. (b) The observed $Q_{1}(2)$ and $^QR_{12}(2)$ lines of \ce{^40CaF} and the fitted curve.}

    \label{fig:fit}
\end{figure}

\begin{figure}[h]
    \begin{subfigure}[b]{0.5\textwidth}
    \centering
        \includegraphics[width=\textwidth]{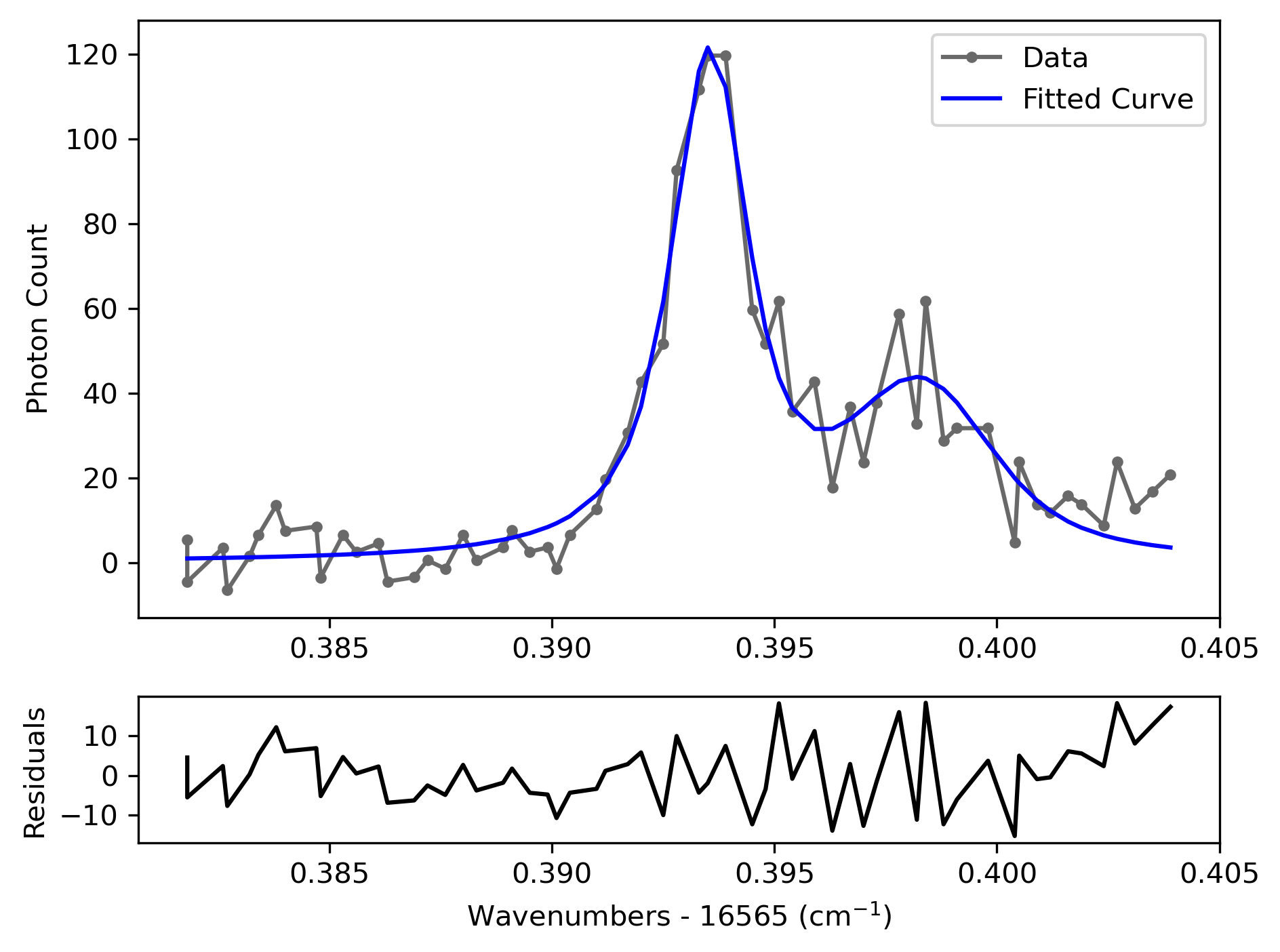}
        \caption{}
    \end{subfigure}
    \begin{subfigure}[b]{0.5\textwidth}
        \centering
        \includegraphics[width=\textwidth]{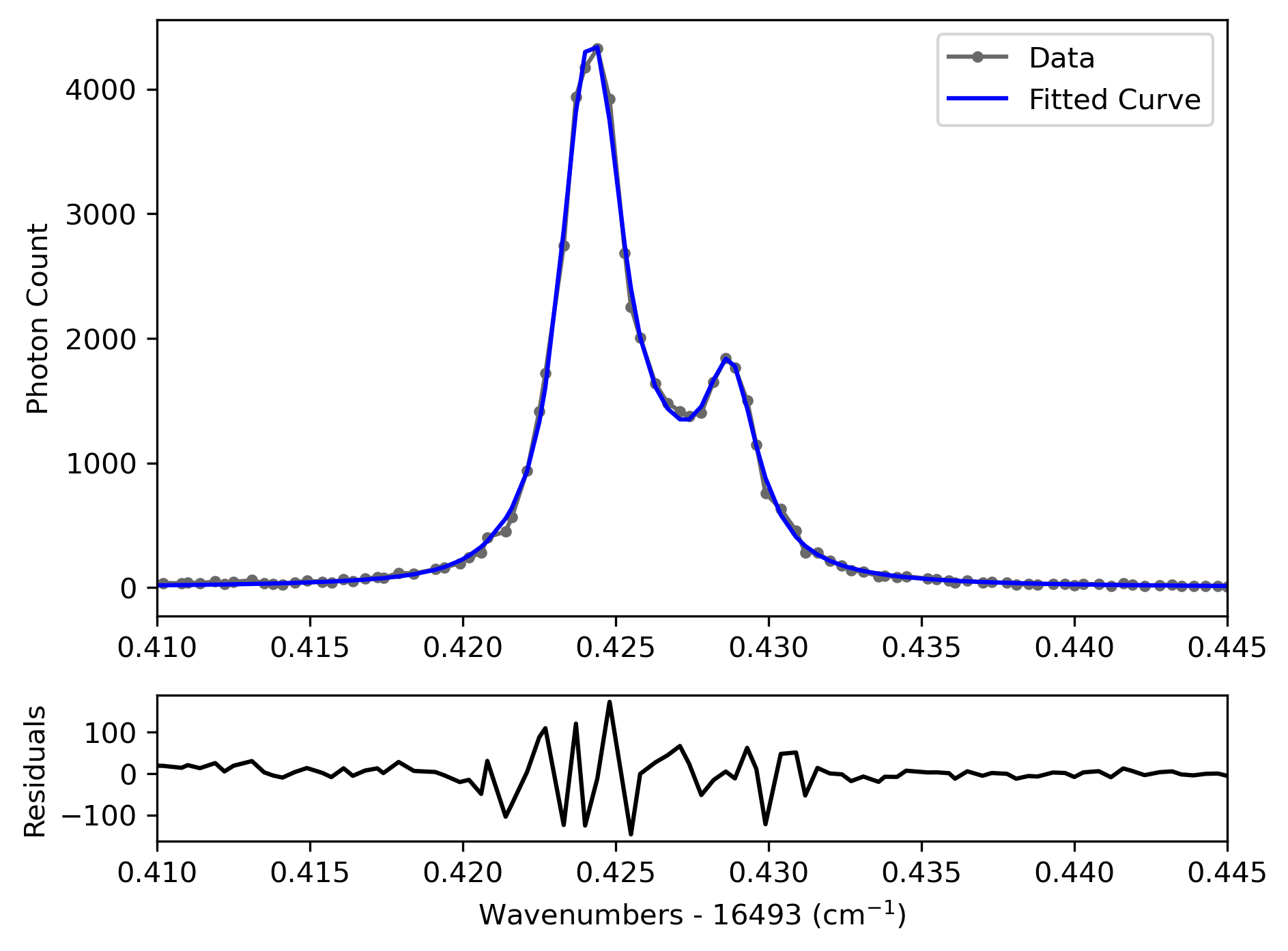}
        \caption{}
    \end{subfigure}
    \caption{(a) The observed $R_2(1)$ and $^RQ_{21}(1)$ lines of \ce{^44CaF} and the fitted curve. (b) The observed $R_2(1)$ and $^RQ_{21}(1)$ lines of \ce{^40CaF} and the fitted curve.}

    \label{fig:fit}
\end{figure}